\documentclass[twocolumn, switch]{article} 

\usepackage{preprint}

\usepackage[utf8]{inputenc}	
\usepackage[T1]{fontenc}	
\usepackage{amsmath, amsthm, amssymb, amsfonts, mathrsfs}
\usepackage{algorithm}
\usepackage{algorithmicx}
\usepackage{algpseudocode}
\usepackage{xcolor}		
\usepackage{booktabs} 		
\usepackage{nicefrac}		
\usepackage{microtype}		
\usepackage[switch]{lineno}	
\usepackage[colorlinks = true,
linkcolor = purple,
urlcolor  = blue,
citecolor = cyan,
anchorcolor = black]{hyperref}	
\usepackage{array}
\usepackage{multirow}
\usepackage{textcomp}
\usepackage{stfloats}
\usepackage{url}
\usepackage{verbatim}
\usepackage{graphicx}
\usepackage{graphics}
\usepackage{lipsum}
\usepackage{subcaption}
\usepackage{xcolor}
\usepackage{soul}
\usepackage{tikz,hyperref}

\usepackage[numbers,square]{natbib}
\usepackage{titlesec}
\titlespacing\section{0pt}{12pt plus 3pt minus 3pt}{1pt plus 1pt minus 1pt}
\titlespacing\subsection{0pt}{10pt plus 3pt minus 3pt}{1pt plus 1pt minus 1pt}
\titlespacing\subsubsection{0pt}{8pt plus 3pt minus 3pt}{1pt plus 1pt minus 1pt}

\definecolor{lime}{HTML}{A6CE39}
\DeclareRobustCommand{\orcidicon}{
	\begin{tikzpicture}
		\draw[lime, fill=lime] (0,0) 
		circle [radius=0.16] 
		node[white] {{\fontfamily{qag}\selectfont \tiny ID}};
		\draw[white, fill=white] (-0.0625,0.095) 
		circle [radius=0.007];
	\end{tikzpicture}
	\hspace{-2mm}
}
\foreach \x in {A, ..., Z}{\expandafter\xdef\csname orcid\x\endcsname{\noexpand\href{https://orcid.org/\csname orcidauthor\x\endcsname}
		{\noexpand\orcidicon}}
}

\title{Actuator-Aware Spatiotemporal Tube Synthesis for Temporal Reach-Avoid-Stay Tasks}

\usepackage{xwatermark}
\newwatermark[firstpage,color=gray!90,angle=0,scale=0.28, xpos=0in,ypos=-5in]{*correspondence: \texttt{keshabpatra19@gmail.com}}

\usepackage{authblk}

\author[1\thanks{\tt{keshabpatra19@gmail.com}}]{Keshab Patra\orcidA{}}
\author[1]{K Madhava Krishna\orcidB{}}

\affil[1]{Robotics Research Center,
	International Institute of Information Technology - Hyderabad
	Hyderabad, Telangana, India}

\begin{document}

\twocolumn[ 
  \begin{@twocolumnfalse} 
  
\maketitle

\begin{abstract}
This work proposes an actuator-aware spatiotemporal tube (STT) synthesis framework to accomplish temporal reach-avoid-stay (T-RAS) tasks for an unknown nonlinear multi-input and multi-output (MIMO) system under actuator constraints. Existing STT synthesis methods address actuator saturation after the tube generation either through repeated online re-optimization or controller redesign. Instead, the proposed framework incorporates actuator constraints directly into the tube synthesis process. The STT centerline and width are parameterized using Bernstein polynomial basis functions, whose convex-hull property enables sample-free enforcement of geometric and derivative constraints. By analyzing the worst-case closed-loop error
dynamics of an approximation-free prescribed performance controller (PPC) used for STT tracking, we derive a linear actuator feasibility constraint. The constraints are embedded directly in terms of tubes' Bernstein control points into the STT synthesis
optimization for actuator-feasible tube generation, eliminating the need for online re-optimization or controller redesign. A simulation study on an omnidirectional mobile robot performing a T-RAS task shows that the proposed framework adheres to the prescribed actuator limits throughout the task and reduces required control effort by approximately 50\% compared with an existing STT synthesis method.
\end{abstract}
\vspace{0.35cm}

  \end{@twocolumnfalse} 
] 



\section{Introduction}

An autonomous system, like a mobile robot navigating through a sandy ground along with a noisy obstacle map, an unmanned aerial vehicle (UAV) flying at high speed through a forest in the presence of wind gusts, a legged robot traversing through rough terrain, or a mobile manipulator grasping and manipulating an object in an environment with uncertainty. Such applications require the robot to start from an initially specified location and reach desired targets with high agility while avoiding collisions with proximity, in compliance with its admissible state and control bounds. Such a task of reaching a target safely while avoiding obstacles is known as the temporal reach-avoid-stay (T-RAS) specification. Completing tasks becomes challenging when the system dynamics are nonlinear, unknown, or partially known and subject to actuator limits.

Classical path planning algorithms \cite{2023_liu_review}, graph-based methods include A* and Dijkstra, sampling-based methods include PRM, and RRT generates feasible paths, which require downstream tracking control, do not provide time, system constraint requirements, and formal guarantees. Model Predictive Control (MPC) \cite{2021_hang}, like an optimization-based approach, requires a dynamic model. Funnel-based feedback control techniques execute trajectory tracking within a guaranteed bound \cite{2008_Bechlioulis}, are effective for reachability and tracking \cite{2014_Bechlioulis}, and are challenging for the collision avoidance task \cite{2021_Lindemann}. Path planning has been integrated with the funnel-based tracking \cite{2018_Ravanbakhsh} within an obstacle-free space, which requires a trajectory prior from an external planner and does not scale well with the number of obstacles.

The STT framework \cite{2026_das_spatiotemporal} provides a safe time-varying, goal-reaching tube in the state space for an unknown nonlinear control-affine system, without the need for explicit trajectory planning. The core idea is to construct STT functions in the output space in place of the trajectory and use a PPC-style tracking controller to confine the output within the tube at all times. The similar formulation along several directions are Signal Temporal Logic specifications recast as tube constraints~\cite{2025_das_stl}, multi-agent coordination with inter-agent collision avoidance~\cite{2026_basu_multiagent}. The STT requirements were posed as a Robust Optimization Problem (ROP) over an infinite family of continuous-time constraints, and then relaxed to a tractable Scenario Optimization Problem (SOP) via discrete time sampling.

 The resulting abrupt adjustments to the tube shape due to unsafe-set avoidance translate into large, sometimes discontinuous control efforts, motivating a smoothed variant of the circumvent mechanism for static obstacles~\cite{2025_upadhyay_smooth}. However, none of these works couples the tube-shaping to the actuator limit or extends the smoothness treatment to time-varying obstacles or to the tube parametrization. The existing STT and funnel-control frameworks guarantee T-RAS performance by allowing the controller's gain to grow unboundedly, considering unlimited actuation capacity.

The existing work addresses the input constraints \emph{after} tube synthesis as a decoupled problem, either by modifying the controller online in response to saturation events \cite{hopfe2010funnel,berger2022inputconstrained} or by replacing the closed-form control law with a repeatedly solving an optimization problem \cite{berger2022feasibility,berger2024arbitrary,berger2023robust}.

This work is the first, to the best of our knowledge, to propose an actuator-aware STT synthesis framework. We derive an almost sample-free, control input aware tube design exploiting the Bernstein basis function ~\cite{farouki2012bernstein} to parametrize STTs such that PPC controller's actuator demand is bounded by construction, without online re-optimization or controller modifications.

The contributions of this work are summarized:
\begin{enumerate}
  \item We propose an actuator-aware STT synthesis framework for T-RAS tasks of an unknown nonlinear MIMO system. The actuator limits are incorporated directly in the tube design rather than handled via online re-optimization or controller redesign.
  \item We derive an actuator-feasibility condition by analyzing the closed-loop error dynamics. The resulting condition depends only on the geometric properties of the synthesized tube, expressed as a linear constraint on the Bernstein control points enables sample-free actuator-aware STT synthesis.

  \item We develop an approximation-free PPC controller for tracking the synthesized STT. A simulation study of an omnidirectional mobile robot demonstrates that the proposed actuator-aware synthesis significantly reduces control effort compared with existing STT synthesis methods while safely accomplishing the T-RAS task within the prescribed actuator limits.
\end{enumerate}
\section{Preliminaries}
\subsection{System Definition}
A class of $N$-th order control-affine MIMO nonlinear systems with $m$ inputs and $m$ outputs is defined as
\begin{equation}\label{eqn: MIMO_system}
    \begin{aligned}
        \dot{x}_i(t) = f_i(x(t)) + \mathbf{G}_i(x(t))x_{i+1}(t) + \mathrm{d}_i(t),\\
        i\in [1, N-1] \\
        \dot{x}_{N} = f_N(x(t)) + \mathbf{G}_N(x(t))u(t) + \mathrm{d}_N(t)\\
        y(t) = x_1(t)
    \end{aligned}
\end{equation}

where $t \in\mathbb{R}^{+},\ \text{for}\ i \in[1,\cdots,N]$\\
$x_i(t) = [x_{i,1}(t), \cdots, x_{i,m}(t)]^T\in \boldsymbol{\mathrm{X}_{i}}\subset \mathbb{R}^m$, is the state vector,
$x(t) = [x^T_1(t), \cdots, x^T_N(t)]^T\ \in \boldsymbol{\mathrm{X}}\subset\mathbb{R}^{N\cdot m}$ is the state space
$\mathrm{d}_i(t) \in\mathbb{R}^m$ is a bounded unknown external disturbance, $|\mathrm{d}_i(t)| \leq \overline{\mathrm{d}}_i$, $u(t) \in \mathbb{R}^m$ is the control input vector of the system, where as $u_i(t) \forall i \in [1,N-1]$ are the virtual control inputs.
$y(t) = [x_{1,1}(t), \cdots, x_{1,m}(t)]^T \in\boldsymbol{\mathrm{Y}}\subset\mathbb{R}^m\ $ indicates the output vector in the output space $\boldsymbol{\mathrm{Y}}\subset\mathbb{R}^m$.\\
The functions $f_i: \mathbb{R}^m \rightarrow \mathbb{R}^m$ is an unknown smooth nonlinear map, $\ g_{i,j}: \mathbb{R}^{m} \rightarrow \mathbb{R}$ is an unknown control co-efficient and $\mathbf{G}_i := [g^T_{i,1}(x(t)),\ \cdots,\ g^T_{i,m}(x(t))]^T$ \\

\textit{Assumptions 1:} The vector function $f_i$ and the matrix $\mathbf{G}_i$ are unknown, uniformly bounded, continuous with time, and locally Lipschitz for all $i\in [1,N]$. $\underline{f}_i \leq f_i(x_t) \leq \overline{f}_i$\\
\textit{Assumptions 2:} The matrix $\mathbf{G}_i$ is positive definite matrix for all $(x,t) \in \mathbb{R}^m \times \mathbb{R}^+$, i.e. there exist a $\underline{g}_i \in\mathbb{R}^+$ such that $0< \underline{g}_i < \lambda_{min}(\mathbf{G}_i(x(t)), \forall (x,t)\in \mathbb{R}^m \times \mathbb{R}^+$. $\lambda_{min}(\cdot)$ indicates the smallest eigenvalue of the matrix.

The controller tracks the desired output $y_d$, and the tracking error is defined as $z_1 = x_1 - y_d$. We employ a back-stepping control technique, which consists of $N$ error vector as follows
\begin{equation}\label{eqn:tracking_error}
    z_i = x_i - u_{i-1}, \forall i \in [1,\cdots, N]
\end{equation}
where $u_0 = y_d,\ z_i \forall i \in [1,N]$ are error vectors and $z_1$ is an actual output tracking error. $u_i\ \forall i \in [1,N-1]$ are virtual control inputs and $u_n = u$ is the actual control input.  
\subsection{Problem Formulation}
We define the obstacle or the unsafe region as a time-varying set $\boldsymbol{\mathrm{O}} = \prod_{k\in[1, N_o]} \mathcal{O}_k$ that includes all $N_o$ number of obstacles and unsafe regions inside the output space $\boldsymbol{\mathrm{O}}\subset \boldsymbol{\mathrm{Y}}$. The unsafe region $\boldsymbol{\mathrm{O}}$ is compact and can be nonconvex and disjoint.

\textbf{\textit{T-RAS Task Specification:}} For a T-RAS task of an autonomous system is to reach a defined goal set $\boldsymbol{\mathrm{T}} \subset (\boldsymbol{\mathrm{Y}} \setminus \boldsymbol{\mathrm{O}})$ from an initial location $y(0) \in \boldsymbol{\mathrm{S}}\subset (\boldsymbol{\mathrm{Y}} \setminus \boldsymbol{\mathrm{O}}) $ inside a start set $\boldsymbol{\mathrm{S}}$ at a prescribed time $t_g \in \mathbb{R}^+$ while avoiding the unsafe region $y(t) \cap \boldsymbol{\mathrm{O}} = \emptyset\ \forall t \in [0,t_g]$.

\textbf{\textit{Problem}}: For a nonlinear system described in \eqref{eqn: MIMO_system}, we need to design an  approximation-free, closed-form control law $u(t)$ that remains within the admissible limits and ensures that $y(t)$ follows the desired \textit{T-RAS Task Specification}.

\textbf{\textit{STTs for T-RAS Task:}}
For a specified \textit{T-RAS} task, we define STTs as a time varying interval $[\underline{\rho}(t), \overline{\rho}(t)]$ which is parameterized as nominal center trajectory $\mathrm{c}(t)\ \text{with width}\ \mathrm{w}(t)$, where $\underline{\rho}(t) =c(t) - \mathrm{w}(t),\ \overline{\rho}(t) = \mathrm{c}(t) + \mathrm{w}(t)$. A user-defined control scheme $u(t)$ within the actuator limits, constrains the output trajectory $y(t)$ within the STTs $\mathcal{T}(t)$. The STTs for T-RAS task are formally defined as follows:
\begin{equation}\label{eqn:stts_conditions}
\begin{aligned}
       \mathcal{T}(t) = \{ y(t) \in \mathbb{R}^m: |y(t) - \mathrm{c}(t)| \leq\mathrm{w}(t) \} \\
       \text{or}\ \underline{\rho}(t)  \leq y(t) \leq \overline{\rho}(t)\\
\end{aligned}
\end{equation}
where $\mathrm{c}(t): \mathbb{R}^+ \rightarrow \mathbb{R}^m,\ \mathrm{w}(t) \geq \underline{\mathrm{w}} > 0\  \text{and}\ \mathrm{w}(t): \mathbb{R}^+ \rightarrow \mathbb{R}^m$ are continuously differentiable vector valued functions.


%
The STTs must hold the following T-RAS condition
\begin{equation}\label{eqn:t-ras}
    \begin{aligned}
        \quad &\mathcal{T}(0) \subseteq \boldsymbol{\mathrm{S}},\ \mathcal{T}(t_g) \subseteq \boldsymbol{\mathrm{T}}\\
       \quad & \mathcal{T}(t) \subseteq (\boldsymbol{\mathrm{Y}} \setminus \boldsymbol{\mathrm{O}})\ \forall t \in [0,t_g]
    \end{aligned}
\end{equation}

\textbf{\textit{Objective:}} The main objective is to synthesize controller aware smooth STT $\mathcal{T}$ function for the autonomous system described in \eqref{eqn: MIMO_system} which can reach the target region $\boldsymbol{\mathrm{T}}$ safely without reaching the actuator saturation.
\section{Actuator Aware STTs Synthesis}
We construct the STTs as a minimum and a maximum bound along a centerline trajectory $\boldsymbol{\mathrm{c}}(t)$. The half-width of the bound is represented by  $\boldsymbol{\mathrm{w}}(t)$. Both the centerline and the width are represented using Bernstein polynomial basis vectors for the following advantages: i) interpolation between start and goal, ii). the entire trajectory remains within the convex hull of its control points, iii) the derivative of the function is also Bernstein, which allows computation of derivatives from the control points, and iv) preserves closed-form derivative bounds.  

\subsection{STTs Synthesis}
The Bernstein polynomials of degree $d$ form the basis for the STTs. We use the same polynomial degree for the centerline trajectory $\mathrm{c}(t)$ and the width $\mathrm{w}(t)$, and compute the control point vectors $\boldsymbol{\mathrm{C}}$ and $\boldsymbol{\mathrm{W}}$ during synthesis. The representation of the STTs trajectories, $\boldsymbol{\mathrm{c}}(t)$ and $\boldsymbol{\mathrm{w}}(t)$ are in the following.
\begin{subequations}\label{eqn:stts}
    \begin{align}
    \boldsymbol{\mathrm{c}}(t) & = \sum_{i=0}^{d} \boldsymbol{\mathrm{C}}_{i}\boldsymbol{\mathrm{B}}^{d}_i(\tau)\label{eqn:stts_centerline}\\
    \boldsymbol{\mathrm{w}}(t) & = \sum_{i=0}^{d} \boldsymbol{\mathrm{W}}_{i}\boldsymbol{\mathrm{B}}^{d}_i(\tau) \label{eqn:stts_width}\\
    &\text{where},\ \tau = t/t_c \in [0,1] \notag
    \end{align}
\end{subequations}
where $\boldsymbol{\mathrm{C}}_i,\ \boldsymbol{\mathrm{W}}_i$ are the unknown $i-th$ control point vector of the STTs, $\boldsymbol{\mathrm{B}}^{d}_i(\tau)$ is the $i$-th basis vector of degree $d$ for the centerline and the width of the STTs, $\tau$ is the parameterized time and $t_c$ is the total trajectory time.
The STTs derivative $\dot{\boldsymbol{\mathrm{c}}}(t)$ and $\dot{\boldsymbol{\mathrm{w}}}(t)$ are
\begin{subequations}\label{eqn:stts_derivative}
    \begin{align}
    \dot{\boldsymbol{\mathrm{c}}}(t) & = \frac{d}{t_c}\sum_{i=0}^{d-1} (\boldsymbol{\mathrm{C}}_{i+1} - \boldsymbol{\mathrm{C}}_{i}) \boldsymbol{\mathrm{B}}^{d-1}_i(\tau) \label{eqn:stts_centerline_vel}\\
    \dot{\boldsymbol{\mathrm{w}}}(t) & =  \frac{d}{t_c} \sum_{i=0}^{d-1} (\boldsymbol{\mathrm{W}}_{i+1} - \boldsymbol{\mathrm{W}}_{i}) \boldsymbol{\mathrm{B}}^{d-1}_i(\tau) \label{eqn:stts_width_rate}
    \end{align}
\end{subequations}

We formulate the STTs synthesis problem as a nonlinear optimization problem (NLP) that satisfies the STTs constrained in \eqref{eqn:stts_conditions} and \eqref{eqn:t-ras}. The NLP formulation is described as follows.
\begin{subequations}\label{eqn:stts_opt}
    \begin{align}
        \boldsymbol{\mathrm{C}}^*, & \boldsymbol{\mathrm{W}}^*\ =\ \min_{\boldsymbol{\mathrm{C}}, \boldsymbol{\mathrm{W}}} \sum_{i=0}^{d-1} J_i \label{eqn:stt_opt_cost_function} \\
\text{s.t.} \quad & \boldsymbol{\mathrm{W}}_{i} \geq \boldsymbol{\mathrm{w}}_{min},\ \forall i \in [1,d],\  \boldsymbol{\mathrm{w}}_{min} > 0 \label{eqn:stt_opt_tube_width_constr}\\ 
        & \frac{d}{t_c}(\boldsymbol{\mathrm{C}}_{i+1} - \boldsymbol{\mathrm{C}}_{i}) \leq \dot{\boldsymbol{\mathrm{c}}}_{max},\ \forall i\in [0,d-1]\label{eqn:stt_opt_center_traj_vel_constr}\\ 
        & \frac{d}{t_c}(\boldsymbol{\mathrm{W}}_{i+1} - \boldsymbol{\mathrm{W}}_{i}) \geq \dot{\boldsymbol{\mathrm{w}}}_{min},\ \forall i\in [0,d-1] \label{eqn:stt_opt_tube_width_rate_contraction_constr}\\ 
        & (\dot{\boldsymbol{\mathrm{c}}}_{max} - \dot{\boldsymbol{\mathrm{w}}}_{min}) \leq (\alpha * \overline{u}_1 - \beta) \label{eqn:stt_opt_actuator_constraints}\\
        & ||\boldsymbol{\mathrm{C}}_{0} - y(0)|| \leq \delta \label{eqn:stt_opt_ic}\\ 
        & ||\boldsymbol{\mathrm{C}}_{d} - y(t_c)|| + \boldsymbol{\mathrm{w}}(t_c) \leq \rho_{\mathrm{G}} \label{eqn:stt_opt_tc}\\ 
        & \mathcal{T}(t) \cap \mathcal{O}_j = \emptyset,\ \forall j\in [1, N_{o}]\label{eqn:stt_opt_coll_constr} 
    \end{align}
\end{subequations}
where the subscript $i$ refers to the control point, $J_i$ in \eqref{eqn:stt_opt_cost_function} is the cost function (Section \ref{sec:cost_function}), $u_{max}$ is the control limit and  $\alpha,\ \beta\ $ are derived from the controller design in Section \ref{sec:actuator_aware_controller}, $\delta$ is a positive constant. Eqn.\eqref{eqn:stt_opt_tube_width_constr} ensures positive tube width, \eqref{eqn:stt_opt_center_traj_vel_constr} ensures the velocity of centerline trajectory remain within a variable bound $\dot{\boldsymbol{\mathrm{c}}}_{max}$, \eqref{eqn:stt_opt_tube_width_rate_contraction_constr} controls the tube width contraction rate $\dot{\boldsymbol{\mathrm{w}}}_{min}$, \eqref{eqn:stt_opt_actuator_constraints} ensure control aware STTs to avoid control limit saturation, \eqref{eqn:stt_opt_ic} and \eqref{eqn:stt_opt_tc} are the terminal constraints and \eqref{eqn:stt_opt_coll_constr} ensures collision avoidance. The cost function, tube constraints, terminal constraints, and obstacle avoidance constraints are described in detail in Sections \ref{sec:cost_function}, \ref{sec:tube_constraints}, and \ref{sec:obstacle_avoidance}, respectively.

\subsubsection{Cost Function} \label{sec:cost_function}
The objective of the STTs synthesis optimization in \eqref{eqn:stts_opt} is to find a feasible tube trajectory with minimum centerline velocity $\dot{c}(t)$ that helps in minimizing the control efforts. We include the variable terms of the centerline velocity described in \eqref{eqn:stts_centerline_vel}. The cost function is as follows
\begin{equation}\label{eqn:stts_opt_cost_func_desc}
    J_i = ||\boldsymbol{\mathrm{C}}_{i+1} - \boldsymbol{\mathrm{C}}_{i}||^2
\end{equation}

\subsubsection{Tube Constraints} \label{sec:tube_constraints}
The tube constrained \eqref{eqn:stt_opt_tube_width_constr}, \eqref{eqn:stt_opt_center_traj_vel_constr}, and \eqref{eqn:stt_opt_tube_width_rate_contraction_constr} imposed on the STTs synthesis \eqref{eqn:stts_opt} to meet the T-RAS task. Eqn. \eqref{eqn:stt_opt_tube_width_constr} ensures a positive finite tube width where $\mathrm{w}_{min}$ incorporates the robot geometry and ensures that minimum error bounds are met within the actuator limits. The maximum of tube centerline velocity is limited by a variable rate limit constraint $\dot{\boldsymbol{\mathrm{c}}}_{max}$ presented in \eqref{eqn:stt_opt_center_traj_vel_constr}. The tube width contraction rate is limited by a variable rate limiter $\dot{\boldsymbol{\mathrm{w}}}_{min}$. The actuator aware tube synthesis is formulated as a constraints \eqref{eqn:stt_opt_actuator_constraints} on the variable centerline velocity bound $\dot{\boldsymbol{\mathrm{c}}}_{max}$ and tube width contraction $\dot{\boldsymbol{\mathrm{w}}}_{min}$ limits which has a linear relationship with the actuator limit $u_{max}$, which we compute from the actuator-aware control design illustrated in Section \ref{sec:actuator_aware_controller}.

\subsubsection{Terminal Constraints} \label{sec:terminal_constraints}
To complete the T-RAS task, the STTs must start from the start region. The initial centerline trajectory $\boldsymbol{\mathrm{c}}(0)$ or $\boldsymbol{\mathrm{C}}(0)$ should starts from the specified location $y(0)$ within some positive bound $\delta$. The initial condition has been imposed on the \eqref{eqn:stt_opt_ic}. The goal-reaching condition is imposed in \eqref{eqn:stt_opt_tc}, ensuring that the robot reaches the goal set at the prescribed time $t_c$.

\subsubsection{Obstacle Avoidance} \label{sec:obstacle_avoidance}
The obstacles $\mathcal{O} \in \boldsymbol{\mathrm{O}}$ are considered convex shapes or substituted with a union of multiple convex shapes for nonconvex obstacles. A set of inequality constraints represents an obstacle $\mathcal{O}$
\begin{equation}
    \mathcal{O} = \{\boldsymbol{\mathrm{x}} \in \mathbb{R}^{n_d}: \boldsymbol{\mathrm{A}}\boldsymbol{\mathrm{x}} \leq \boldsymbol{\mathrm{b}}, \boldsymbol{\mathrm{A}} \in \mathbb{R}^{n_f\times n_d},\ \boldsymbol{\mathrm{b}} \in \mathbb{R}^{n_f}\}
\end{equation}
where $n_d$ and $n_f$ are the dimension and the number of faces of convex obstacle $\mathcal{O}$ and $\boldsymbol{\mathrm{x}}$ is an interior point.

The obstacle avoidance constrained are defined as
\begin{equation} \label{eqn:stt_obs_constr}
\begin{aligned}
    & \boldsymbol{\mathrm{p}}_j =  \boldsymbol{\mathrm{A}}_j(\boldsymbol{\mathrm{c}}(t) \pm \boldsymbol{\mathrm{w}}(t)) - \boldsymbol{\mathrm{b}}_j - d_{safe}\\
   & \frac{1}{k}\log\left [{\sum \exp({k*\boldsymbol{\mathrm{p}}_j}})\right] \geq 0\\
   & \forall j \in [1, N_o]\ \text{and}\ \forall t
\end{aligned}
\end{equation}
where $k$ is a positive constraints typical value $k>10$
The STTs should remain outside $\boldsymbol{\mathrm{O}}$, \eqref{eqn:stt_obs_constr} ensures the collision avoidance constraints \eqref{eqn:stt_opt_coll_constr}.

\subsection{Actuator-aware Controller Design} \label{sec:actuator_aware_controller}
We use the STTs defined in \eqref{eqn:stts} to design an approximation-free closed-form control law based on prescribed performance control. The lower-triangular structure permits a back-stepping like controller design approach, similar to \cite{2014_Bechlioulis}.

We design a control input $u_1$ that ensures $x_1$ tracks the desired trajectory $y_d(t)$ and follows \eqref{eqn:stts_conditions}. We then iteratively design a control law $u_i$ for the intermediate dynamics $x_i$ of the system (\ref{eqn: MIMO_system}) which ensures $x_i$ tracks $u_{k-1},\ \forall i \in [2,N]$.

The time-varying bounds on the tracking errors $z_i$ are represented by $\phi$. The constraints $\phi_1$ on the desired trajectory tracking error $z_1$ are defined by STTs $\phi_1 = \boldsymbol{\mathrm{w}}(t)$ synthesized in \eqref{eqn:stts_width}. The constraints $\phi_i= [\phi_{i,1},\ \cdots,\ \phi_{i,m}],\ \forall (i,j) \in [2,N] \times [1,m]$ are defined in \eqref{eqn:control_constr}.
\begin{equation}\label{eqn:control_constr}
    \phi_{i,j} = (p_{i,j}-q_{i,j})e^{-\mu_{i,j}} + q_{i,j}
\end{equation}

We normalized the error $z_i$ as $\boldsymbol{e}_i = [e_{i,1},\cdots, e_{i,m}]^T$ described in \eqref{eqn:error_norm}.
\begin{equation} \label{eqn:error_norm}
    e_{i,j} = \frac{z_{i,j}}{\phi_{i,j}}\ \forall (i,j) \in [1,N]\times[1,m]
\end{equation}

The control inputs are designed as
\begin{equation}\label{eqn:control_scheme}
\begin{aligned}
      u_i = -\overline{u}_i \frac{2}{\pi}\arctan\left(\boldsymbol{K_i} \ln{\frac{1-\boldsymbol{\mathrm{e}}_i}{1 + \boldsymbol{\mathrm{e}}_i}}\right)
\end{aligned}
\end{equation}
where $\boldsymbol{K_i} := diag([k_1, \cdots, k_m])$ is a positive definite and diagonal matrix.

Taking the derivative of \eqref{eqn:control_scheme}, we get
\begin{subequations}\label{eqn:control_rate}
\begin{align}
    \dot{u}_i &= \psi_i \dot{e}_{i},
    \label{eqn:control_rate_a}\\
    \dot{\boldsymbol{e}}_{i} &=
    \frac{\dot{z}_i\phi_i - z_i\dot{\phi}_i}{\phi_i^2},
    \label{eqn:control_rate_b} \forall i \in [1,N]\\
    \psi_i &=
    \frac{d u_i}{d\boldsymbol{e}_i}
    =
    \frac{4 \overline{u}_i \boldsymbol{K}_i\boldsymbol{e}_i}
    {\pi \left [1  + \left(\boldsymbol{K}_i\ln{\frac{1-\boldsymbol{\mathrm{e}}_i}{1 + \boldsymbol{\mathrm{e}}_i}}\right)^2 \right]},
    \label{eqn:control_rate_c}\\
    &  \notag
\end{align}
\end{subequations}
The bound of $\dot{\boldsymbol{e}}_{i}$ and $\psi_i$ can be infer from \eqref{eqn:control_rate_b} and \eqref{eqn:control_rate_c} respectively as follows

\begin{subequations}\label{eqn:control_rate_bounds}
\begin{align}
    |\dot{\boldsymbol{e}}_{i}| \leq \left( \frac{|\dot{z}_i|}{\boldsymbol{\mathrm{q}}_i} + \frac{\mu_i(\boldsymbol{\mathrm{p}}_i - \boldsymbol{\mathrm{q}}_i)}{\boldsymbol{\mathrm{p}}_i} \right)\ \forall i \in[2,N] \label{eqn:control_rate_bounds_a} \\
    |\psi_i| \leq 2 \overline{u}_i \boldsymbol{K}_i,\ \boldsymbol{e}_i \in (-1,1)\ \text{and}\ \forall i \in[1,N] \label{eqn:control_rate_bounds_b} 
\end{align}
\end{subequations}

Differentiating the tracking error $z_i$ defined in \eqref{eqn:tracking_error} and substituting the $\dot{x}_i$ from \eqref{eqn: MIMO_system}

    \begin{equation}
    \begin{aligned}
        \dot{z}_i = f_i + \mathbf{G}_ix_{i+i} + \boldsymbol{\mathrm{d}}_i - \dot{u}_{i-1}
    \end{aligned}
    \end{equation}
    
The worst-case uncertainty bounds, i.e., the outward motion of the system defined in \eqref{eqn: MIMO_system},can be computed by 
\begin{equation}\label{eqn:uncertainty_bound}
\begin{aligned}
        |\dot{z}_i| \leq \overline{f}_i + \overline{g}_i(\overline{u}_i + p_{i+1}) +  \overline{d}_i + |\dot{u}_{i-1}| \ \forall i \in [i,N-1]\\
        |\dot{z}_N| \leq \overline{f}_N + \overline{g}_N\overline{u} +   \overline{d}_N + |\dot{u}_{N-1}|
\end{aligned}
\end{equation}
From \eqref{eqn:control_rate}, the derivative of the $i-1$-th virtual control input is $\dot{u}_{i-1}= \psi_{i-1} \dot{e}_{-1}$ and substituting respective the bounds from \eqref{eqn:control_rate_bounds} the bound $|\dot{u}_{i-1}|\ \forall i \in [2,N]$ is computed as
\begin{equation}\label{eqn:control_rate_bounds_c}
    |\dot{u}_{i-1}| \leq \left( \frac{|\dot{z}_{i-1}|}{\boldsymbol{\mathrm{q}}_{i-1}} + \frac{\mu_i(\boldsymbol{\mathrm{p}}_{i-1} - \boldsymbol{\mathrm{q}}_{i-1})}{\boldsymbol{\mathrm{p}}_{i-1}} \right) |\underline{\psi}_{i-1}|
\end{equation}

Therefore the bound on outward motion of the system is determined as
\begin{subequations} \label{eqn:bound}
    \begin{align}
        \varphi_i = \overline{f}_i + \overline{g}_ip_{i+1}+  \overline{\mathrm{d}}_i + a_{i-1}\ \forall i \in [1,N-1] \label{eqn:bound_stage1}\\ 
        \varphi_N = \overline{f}_N + \overline{\mathrm{d}}_N + a_{N-1} \label{eqn:bound_stageN}
    \end{align}
\end{subequations}
where $a_{0} = \dot{c}_{max}$ and $a_{i} = (\frac{\varphi_i }{q_i} - \frac{\mu_i(q_i - p_i)}{p_i}) |\underline{\psi_i}|,\ \forall i \in [1,N-1]$ substituted from the bounds in \eqref{eqn:control_rate_bounds_c}

The available control for the guaranteed minimum restoring action is $\underline{g}_i \overline{u}_i$ and combining the control term $\overline{g}_i\overline{u}_i$ from \eqref{eqn:uncertainty_bound}, the total guaranteed restoring action from control input is
 $(\overline{g}_i + \underline{g}_i)\overline{u}_i$

The performance constraint $\phi_1 = \boldsymbol{\mathrm{w}}(t)$ is computed considering the input constraint $\overline{u}$ such that the worst-case uncertainty \eqref{eqn:bound} must not exceed the available guaranteed restoring action. The performance specification $\phi_i$ should satisfy the following constraints for the prescribed control input constrained $\overline{u}$:
\begin{subequations}
\begin{align}
    \varphi_1 \leq (\overline{g}_1 + \underline{g}_1)\overline{u}_1 + \inf_{t} \dot{\boldsymbol{\mathrm{w}}} (t)\label{eqn:control_authority_stage1}\\
    \varphi_i \leq (\overline{g}_i + \underline{g}_i) \overline{u}_i + \mu_i (q_i - p_i)\ \forall i \in [2,N] \label{eqn:control_authority_stagei}
\end{align}
\end{subequations}
with initial condition $|z_i(0)| < \phi_i(0)$

The system output and input will follow its performance constraints, i.e. $z_1(t) < \phi_1(t)\ \text{or}\ z_1(t) < \boldsymbol{\mathrm{w}}(t)$ and $|u(t)| < \overline{u}\ \forall t \in \mathbb{R}^+$, respectively, while the STT synthesis satisfies the feasibility conditions in \eqref{eqn:bound_stage1} and \eqref{eqn:control_authority_stage1}. The feasibility condition for the STT synthesis from the evaluation of $\varphi_1$ is as follows
\begin{equation}\label{eqn:stt_derived_actuator_constr}
    \begin{aligned}
        \overline{f}_1 + \overline{g}_1p_{2}+ \overline{d}_1 + \dot{c}_{max} \leq (\overline{g}_1 + \underline{g}_1)\overline{u}_1 + \dot{\boldsymbol{\mathrm{w}}}_{min} \\
        \dot{c}_{max} - \dot{\mathrm{w}}_{min} \leq (\overline{g}_1 + \underline{g}_1)\overline{u}_1 - (\overline{f}_1 + \overline{g}_1p_{2}+ \overline{d}_1)
    \end{aligned}
\end{equation}
where $\dot{\mathrm{c}}_{max} =\sup_{t} \dot{\boldsymbol{\mathrm{c}}}(t)$ and $\dot{\mathrm{w}}_{min} = \inf_{t} \dot{\boldsymbol{\mathrm{w}}} (t)$ are the constraints in STTs. The STT synthesis constrains in \eqref{eqn:stt_opt_actuator_constraints} is now formalized and compared with \eqref{eqn:stt_derived_actuator_constr}, where $\alpha = \overline{g}_1 + \underline{g}_1$ and $\beta = \overline{f}_1 + \overline{g}_1p_{2}+ \overline{d}_1$. The constraints are then linked with the Bernstein control points \cite{farouki2012bernstein}.

\section{Result}
To show the efficacy of the proposed approach of actuator aware STTs construction for T-RAS task, we implement our technique on an omnidirectional ground robot. We compare our approach with the similar STT construction technique proposed by \textit{Das et al.} in \cite{2026_das_spatiotemporal}

\subsection{Omnidirectional Ground Robot} \label{sec:omni results}
The omnidirectional ground robot model is described as follows:
\begin{equation}
    \begin{bmatrix}
        \dot{x}_1(t) \\
        \dot{x}_2(t) \\
        \dot{x}_3(t)
    \end{bmatrix} = 
    \begin{bmatrix}
        \cos{x_3} & -\sin{x_3} & 0\\
        \sin{x_3} & \cos{x_3} & 0\\
        0 & 0 & 1
    \end{bmatrix}
    \begin{bmatrix}
        u_1\\
        u_2\\
        u_3
    \end{bmatrix} + \boldsymbol{\mathrm{d}}(t)
\end{equation}
\begin{figure}
      \centering
      \includegraphics[width=0.458\textwidth]{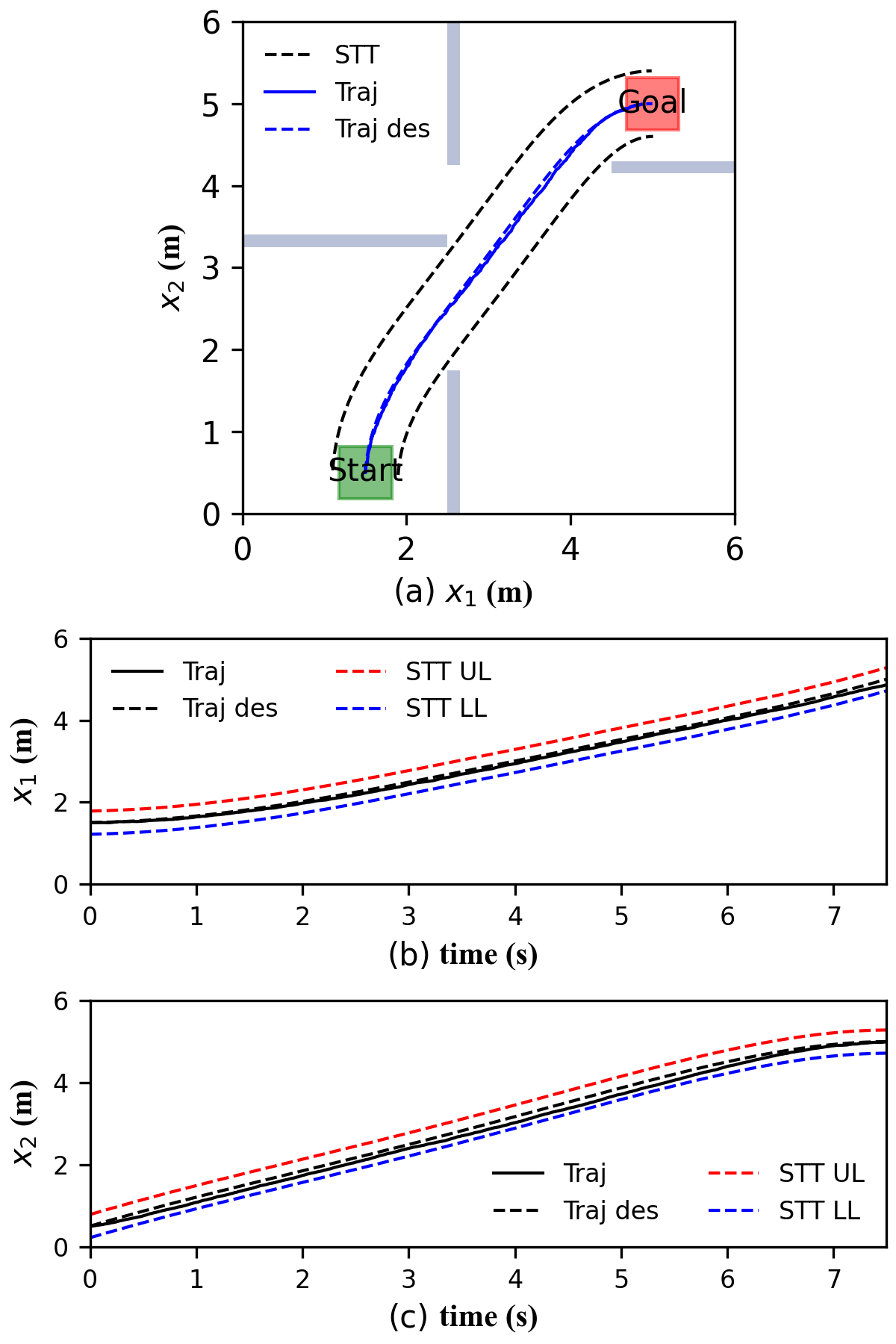}
      \caption{Actuator-aware STT synthesis with STT specification and control input}
      \label{fig:omni task}
   \end{figure}
   
The robot starts from $\boldsymbol{\mathrm{S}} = [-1.1,1.8] \times [0.1, 0.9]$ to reach the goal at $\boldsymbol{\mathrm{T}} =  [4.6, 5.4] \times [4.6, 5.4]$ in $t_c = 7.5\ s$ in a environment with static obstacles shown in Fig. \ref{fig:omni task}(a). We start with a degree time varying $7$ Bernstein polynomial basis function. After solving \eqref{eqn:stts_opt} for the specified task in \eqref{eqn:t-ras}, the values are \\
$\boldsymbol{\mathrm{C}} = [[1.500,1.554,1.928, 2.581, 3.219, 3.694, 4.196,\\ 5.000],[0.500, 1.315, 1.956, 2.528, 3.288, 4.251,  4.988,\\  5.000]]$ 
and 
$\boldsymbol{\mathrm{W}} =[[0.400,0.400, 0.399, 0.400, 0.400,\\ 0.399, 0.401,0.400], [0.400,0.400, 0.399, 0.4000, 0.400,\\ 0.399, 0.401, 0.400]]$

Figure \ref{fig:omni task}(a) shows the executed trajectory by the omnidirectional robot and the STTs are shown in Fig. \ref{fig:omni task}(b).

\subsection{Comparison}
\begin{figure}
      \centering
      \includegraphics[width=0.45\textwidth]{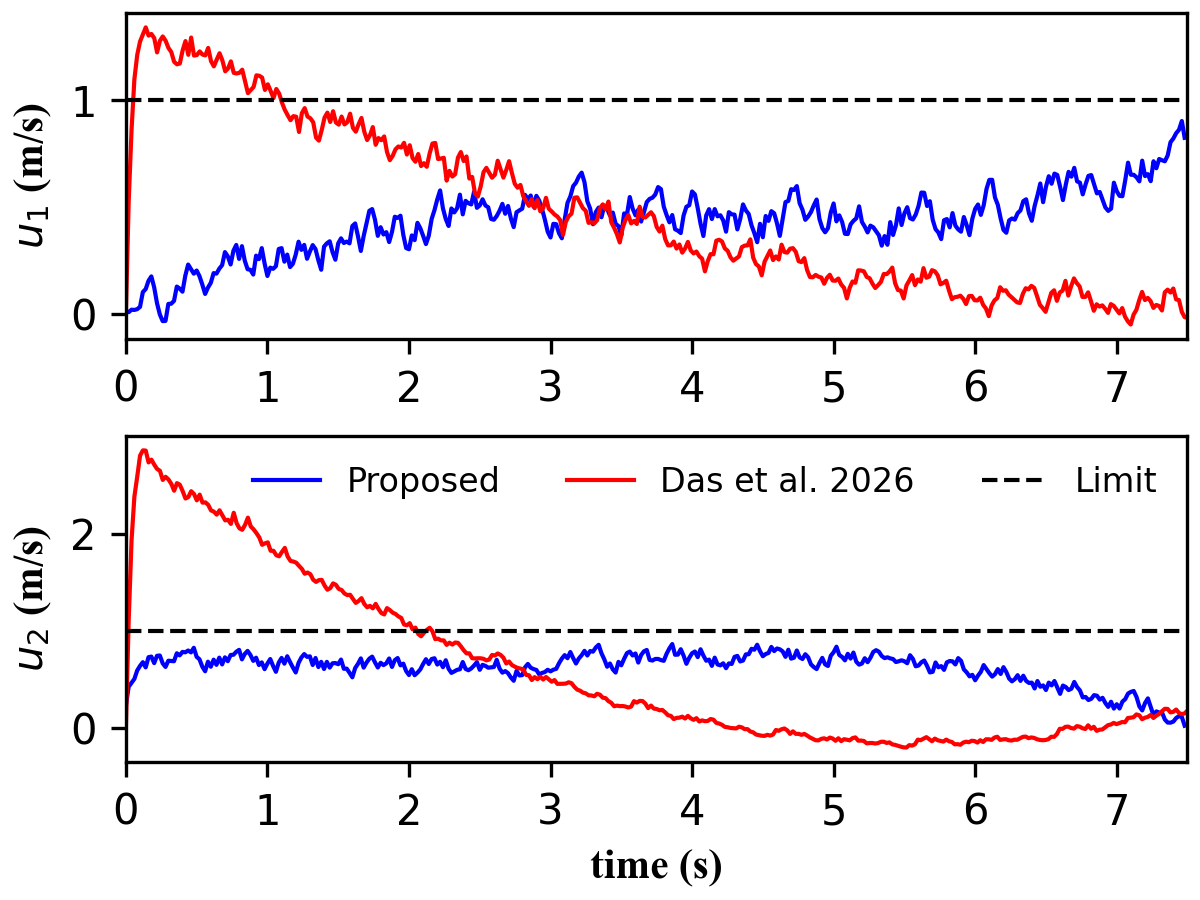}
      \caption{Control effort comparison}
      \label{fig:omni task compare}
\end{figure}

We compare our proposed framework with the STTs design framework available \cite{2026_das_spatiotemporal} for the same omnidirectional robot completing a T-RAS task in the environment in Fig. \ref{fig:omni task}(a). For this comparison we deploy tracking controller without the saturation function $-\overline{u}_i \frac{2}{\pi}\arctan(\cdot)$ used in \eqref{eqn:control_scheme}, i.e. $u = -\boldsymbol{K} \ln{\frac{1-\boldsymbol{\mathrm{e}}}{1 + \boldsymbol{\mathrm{e}}}}$, so that the raw control-effort demand implied by each STT design framework are captured rather than clipped at the actuator limit. The two different STT synthesis frameworks for tube design are compared, while both use the same tracking controller and gain $\boldsymbol{K} = diag([1.5, 1.5])$. Figure \ref{fig:omni task compare} compares the required control input for both methods deployed.
Figure \ref{fig:omni task compare} shows that the required control effort $u_1, u_2$ for the proposed actuator-aware STTs design framework is below the $u_{1,max} = 1.0 m/s$ and $u_{2,max} = 1.0 m/s$ for both $X_1$ and $X_2$ while the required control effort for the STTs design in \cite{2026_das_spatiotemporal} is higher than the threshold. Table \ref{tab:comp} shows the quantitative comparison of the control effor for both the technique.
\renewcommand{\arraystretch}{1.25}
\begin{table}
    \centering
\begin{tabular}{|c|c|c|c|}
        \hline
        \multicolumn{2}{|c|}{}&Proposed & \textit{Das et al.}[7]\\
        \hline
        \multirow{3}{*}{$u_1$}&$u_{1,min}$&0.005&-0.020\\
        \cline{2-4}
        &$u_{1,max}$&0.831&1.273\\
        \cline{2-4}
        &$\sum||u_{1}||^2$&75.690&139.295\\
         \hline
        \multirow{3}{*}{$u_2$}&$u_{2,min}$&0.075&-0.181\\
        \cline{2-4}
        &$u_{2,max}$&0.803&2.724\\
        \cline{2-4}
        &$\sum||u_{2}||^2$&151.134&400.518\\
        \hline
    \end{tabular}
    \caption{Quantitative control effort comparison}
    \label{tab:comp}
\end{table}

\section{Conclusion}
The presented STT synthesis framework generates actuator-aware tubes for T-RAS tasks for unknown nonlinear MIMO systems under prescribed actuator limits. The tube centerline and width parameterization using Bernstein polynomial basis functions, whose convex-hull and derivative properties enable the sample-free formulation of geometric and derivative constraints. The derived actuator-feasibility condition from the prescribed-performance control framework translated into linear constraints on the Bernstein control points. These constraints enable the synthesis of actuator-feasible spatiotemporal tubes without online re-optimization or controller redesign.

Simulation results on an omnidirectional mobile robot demonstrated that the proposed framework successfully completed T-RAS tasks while respecting prescribed actuator limits and significantly reducing the required control effort by $50\%$ compared with existing STT synthesis methods.

Future work will focus on formal stability analysis and recursive feasibility guarantees, extending the formulation to dynamic and uncertain environments with moving obstacles, and experimental validation.
\bibliographystyle{IEEEtran}
\bibliography{ref}
\end{document}